# Yield strength insensitivity in a dual-phase high entropy alloy after prolonged high temperature annealing

Junaid Ahmed      Matthew Daly[*]

*Department of Civil, Materials, and Environmental Engineering, University of Illinois at Chicago – 842 W. Taylor St., 2095 ERF (MC 246), Chicago, IL, 60607, United States*

**ABSTRACT**

Recent studies of FeMnCoCr-based high entropy alloys have demonstrated uncommon deformation behaviors such as transformation-induced plasticity, which were largely believed to be restricted to select families of steels. Coupled with the potential for entropy stabilization of high symmetry phases at high temperatures, this system represents a promising class of materials for structural applications in extreme environments. Yet, transformation-induced plasticity mechanisms are notably sensitive to microstructure parameters and the literature offers examples of deleterious decomposition of high entropy alloys under heat treatment, which raises concerns of resiliency in mechanical performance. Here, we evaluate the evolution of microstructure and mechanical properties of a FeMnCoCr high entropy alloy after prolonged heat treatment at high temperature.  Microstructures are found to retain their characteristic austenite/martensite features, with parent face-centered cubic grains partitioned by hexagonal close-packed laths after heat treatment at 1200 ºC for up to 48 hours. Results of mechanical testing reveal an unusual insensitivity of this alloy to grain growth-induced weakening effects. Namely, the yield strengths of FeMnCoCr samples are observed to remain constant across all heat treatment conditions, despite a near four-fold increase in the grain size. Close examination of post-heat treatment microstructures reveals a dramatic decrease in the inter-lath spacing at longer durations, which segments parent austenite grains. This crystal partitioning counteracts conventional grain growth-induced weakening by introducing additional barriers for dislocation pile-up. These results offer new insights into the mechanical resiliency of this transformation-induced plasticity high entropy alloy under prolonged high temperature heat treatment.

[*]Corresponding author: mattdaly@uic.edu (M. Daly)

---

## 1. INTRODUCTION

Overcoming the traditional tradeoffs between strength and ductility remains a pervasive challenge within the alloy design community. The root of this challenge lies with the prevalence of dislocation slip, which serves as the dominant plasticity mechanism in many engineering metals and alloys. At the mesoscale, the mobility of dislocations mediates two opposing deformation phenomena – microstructure hardening and crystallographic shearing – that underpin the emergence of strength and ductility. In traditional microstructure design, the simultaneous realization of both phenomena represents a non-trivial challenge. For instance, Hall-Petch hardening is a well-known grain boundary strengthening mechanism that describes the inverse relationship between grain size and yield strength [1–4]. Yet, increases to the flow strength through this size effect generally emerge at the expense of ductility. One strategy to circumvent mechanical property tradeoffs is to leverage different deformation mechanisms in alloy design. For instance, twinning-induced plasticity (TWIP) [5] and transformation-induced plasticity (TRIP) [6] mechanisms offer alternative pathways for deformation accommodation in a microstructure. In both mechanisms, deformation is accommodated by a crystallographic reorganization of the microstructure through deformation twinning (i.e., in TWIP) or strain-induced phase transformations (i.e., in TRIP). These alternative deformation pathways simultaneously harden the microstructure through the introduction of new interfaces while also accommodating strain, leading to exceptional combinations of strength and ductility. For example, an early work from Grässel et al. [6] reports an extraordinary combination of ductility (82%) and strength (800 MPa) in a $FeMn_{25}Si_3Al_3$ alloy due to TWIP. On the other hand, Sohrabi et al. [7] reported a combination of excellent ductility (~115%) and strength (~550 MPa) in a TRIP-activated, 304L-series austenitic stainless steel. Indeed, reports of TWIP and TRIP behaviors are a common feature of high performance, technologically-relevant alloys including high Mn steels [8,9], metastable austenitic stainless steels [10,11] and/or TRIP-assisted steels [12–14].

The TRIP/TWIP effects often emerge in closely related alloys that are historically found within select families of the high Mn steels or austenitic stainless steels [15–18], and were therefore largely believed to be relatively rare deformation phenomena. However, the emergence of high entropy alloys (HEA)s [19–21] has opened a wide, and largely unexplored alloy design space that offers an opportunity to re-examine this perception. We note that the community uses several overlapping definitions to refer to these multicomponent systems (e.g., multi-principal element alloys, complex concentrated alloys, or random alloys) [22,23]. As consensus is still emerging on the nomenclature, we use the term high entropy alloys broadly to refer to multicomponent, concentrated systems, which generally aligns with the descriptions of authors in the works referenced herein. While initial reports of the mechanical behavior of single phase HEAs were focused on solid solution strengthening mechanisms [19,24,25] and the prevalence of slip-dominated plasticity [24–26], recent studies report observation of TWIP behavior in several HEA systems including equiatomic $Fe_{20}Mn_{20}Co_{20}Cr_{20}Ni_{20}$ [27] and non-equiatomic $Fe_{40}Mn_{40}Co_{10}Cr_{10}$ [28]. Building on the latter work, Li et al. [29] explored a range of FeMnCoCr compositions and reported a transition in the dominant deformation mechanism from dislocation slip in single-phase (45% Mn) alloy to TWIP in a single phase (40% Mn) alloy to TRIP in a dual-phase (DP) (30% Mn) HEA system. In addition to this DP-HEA, other HEA systems have been reported to exhibit the TRIP effect including Co-rich FeMnCoCrNi [30], quinary FeMnCoCrSi [31], and sexanary FeMnCoCrSiAl [32]. These studies highlight an expansion of TWIP/TRIP mechanisms beyond steels into the realm of HEAs and offer new opportunities to realize these deformation behaviors in this broad alloy design space. However, as HEAs are viewed as a promising material class for high temperature structural applications, the resiliency of TWIP/TRIP microstructures in extreme environments becomes a relevant concern. Indeed, studies from the literature have reported degradation and decomposition of HEA microstructures in a variety of systems [33,34], including the benchmark equiatomic FeMnCoCrNi alloy [35].

Activation of deformation-induced structural transformations such as the TRIP effect is particularly sensitive to microstructural parameters (e.g., volume fraction of retained austenite phase, grain size, and chemical homogeneity) that underpin the relative stability of the austenite phase. Consequently, systems that are optimized to deliver TRIP behavior can be significantly affected by microstructure changes induced by service temperatures [36]. For instance, Samajdar et al. [37] reported a degradation of mechanical properties in TRIP-assisted steel due to a monotonic reduction in the austenite phase at higher heat treatments. Similarly, Lee et al. [38] studied the effect of retained austenite phase fraction on the mechanical properties of TRIP-assisted steel and concluded that the system with higher austenite volume fractions exhibit the best combination of strength and ductility. In addition to phase fraction [39–41], the grain size strongly influences the relative stability of the austenite phase. Several reports have highlighted the link between reductions in the parent austenite grain size and suppression of the martensitic transformation [42–44]. Correlative evidence also underscores the importance of chemical homogeneity amongst constituent phases for realization of the TRIP effect. For example, Li et al. [29] leveraged atom probe tomography to demonstrate a uniform composition along interfaces between austenite and martensite phases of the FeMnCoCr TRIP DP-HEA. Conversely, Wu et al. [2] observed a degradation of TRIP behavior due to precipitation of σ-phase in a $Fe_{30}Mn_{24}Co_{20}Cr_{20}Ni_6$ HEA. Within the context of high temperature applications, the sensitivities of HEA systems and TRIP phenomena underscore the importance for detailed understanding of microstructure evolution in extreme environments, where prolonged heat treatment can manifest in deleterious performance effects such as constituent segregation, microstructure decomposition, and grain growth.

Here, we examine the mechanical properties of the TRIP-assisted, $Fe_{50}Mn_{30}Co_{10}Cr_{10}$ DP-HEA reported by Li et al. [29,45] under prolonged heat treatment at high temperature. Using a combination of microstructure and crystallographic analysis, and mechanical characterization, the effects of heat treatment on the evolution of microstructure and structure-property relationships of this DP-HEA properties are studied. We anticipate that these results will expand the broader application of TRIP-assisted HEAs by exploring the resiliency of their microstructure and mechanical performance in extreme environments.

## 2. MATERIALS AND METHODS

### 2.1. Ingot Casting and Thermomechanical Processing

Beginning with pure elements (purity > 99.9%), an ingot of $Fe_{50}Mn_{30}Co_{10}Cr_{10}$ (at. %) composition was produced by the arc melting method in a vacuum arc melter (VAM) with a water-cooled Cu crucible. Prior to melting, the VAM was evacuated to ~30 mtorr and then backfilled three times with argon (Ar) to purge residual oxygen. After backfilling, the Ar supply was throttled to maintain a constant working pressure of 12 torr (absolute). Upon reaching the working pressure, arc melting proceeded using a direct current welding power supply in the DCEN configuration with a pure tungsten electrode at an operating current of 150 A. The ingot was flipped and re-melted at least 5 times to ensure complete melting and thorough mixing of all constituents. Upon final melting, the ingot was cooled in the crucible under Ar protection. The as-cast ingots were hot-rolled at 900 ºC to reduce the thickness of the specimen, followed by water quenching. The samples were subsequently annealed at 1200 ºC for holding time of 2, 12, 24, and 48 h in an alumina tube furnace followed by water quenching. This heat treatment was selected to explore the effects of extreme temperatures on the microstructure evolution of this HEA system. All heat treatments were performed under an Ar atmosphere. We distinguish each sample here by its final heat treatment time and refer to sample sets as 2h, 12h, 24h and 48h, respectively.

### 2.2. Microstructure and Mechanical Characterization

X-ray diffraction (XRD) analysis was performed using a Bruker D8 Discover diffractometer to study the phases formed in the samples heat treated at different holding times. The XRD analysis was performed over a 2θ range of 30 – 100º in angular increments of 0.05º using a Cu Kα X-ray source ($\lambda = 0.1541$ nm). A JEOL-IT500HR field emission scanning electron microscope equipped with an electron backscatter detector (EBSD) and an energy dispersive X-ray (EDX) spectrometer was used to perform electron microscopy, and crystallographic and chemical analysis. Electron microscopy samples were ion milled using a JEOL SM-09010 cross-sectional polisher at an acceleration voltage of 6 kV. All electron microscopy was performed at acceleration voltages of 15-20 kV and EBSD analysis was executed with a step size of less than 2 μm. EBSD data was post-processed and visualized using the freely available software MTEX [46]. For the visualization of the crystal interfaces, a 10° misorientation criterion was used. For optical microstructural analysis, metallography samples were sequentially ground using SiC papers down to 1200 grit and polished using a 0.05 μm alumina suspension. The microstructure was revealed by immersion etching in a solution comprised of 50 g/L $CuCl_2$ dissolved in equal parts HCl and ethanol (by volume) for 30 s. A Leica DMLM compound microscope was used to perform optical microscopy.

Mechanical characterization consisted of both microhardness indentation testing and uniaxial tensile testing. A Leica Vickers microhardness tester was used to measure the indentation hardness of processed ingots. Indentation measurements were performed under a 200 gf load with a 15 s dwell time and data is reported from at least 10 measurements per condition. Rectangular dog-bone samples (see Figure 1) were machined from the heat-treated specimens and the sample surfaces were mechanically deburred using SiC grinding paper. Dog-bone geometries were specified following the recommendations in Ref. [47] for subsize specimens. Uniaxial tensile tests were performed using a universal testing machine at a strain rate of $10^{-4}$/s. The compliance-corrected strain was measured via digital image correlation (DIC) using the Ncorr software toolbox in MATLAB [48]. For reproducibility, 3 samples at each heat treatment condition were tested at room temperature. Each tensile coupon was machined from a separate ingot in order to assess variability in the castings and materials processing. All error bars are presented as ± 1 standard deviation.

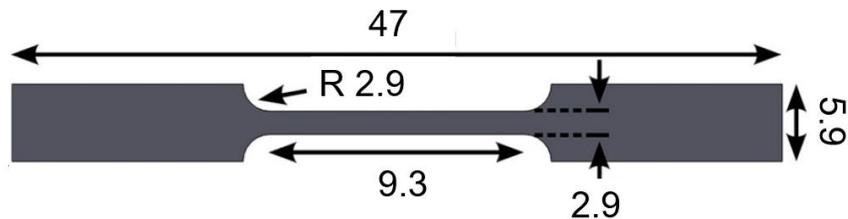

**Figure 1:** Rectangular dog-bone sample used for uniaxial tensile testing. All the dimensions are in mm. See Ref. [47] for further details.

## 3. RESULTS

### 3.1. Microstructure analysis

The XRD spectra collected from the alloys heat treated for 2h and 48h at 1200 ºC are presented in Figure 2. Examination of the XRD data revealed a dual-phase microstructure containing both face-centered cubic (FCC) and hexagonal close-packed (HCP) phases, which is consistent with the phases reported by Li et al. [29] for this TRIP DP-HEA. Based on this XRD data, we measure unit cell parameters of c = 0.411 nm and a = 0.253 nm (c/a = 1.622) and a = 0.356 nm for the HCP and FCC phases, respectively. Similar results are obtained for the other heat treatment conditions and their XRD patterns are provided in the Supplementary Material. Figure 3a presents an optical micrograph of the 2h sample. Examination of the etched surface reveals features consistent with a dual-phase austenite/martensite microstructure. Namely, the parent austenite phase appears to be partitioned by laths of martensite that transect individual grains. Using a series of optical microscopy images, the parent austenite grain size of the 2h sample was measured to be 102.1 ± 11.2 µm from the line intercept method. As anticipated, the grain size was observed to monotonically increase with heat treatment time and reached a size of 404.4 ± 23.2 µm for the 48h condition. The grain sizes for the 12h and 24h samples were measured to be 170.7 ± 15.1 µm and 270.0 ± 19.0 µm, respectively. Optical micrographs for the 12h, 24h, and 48h samples show similar morphologies and are provided in the Supplementary Material.

In order to clearly image microstructural features, high resolution electron channeling-contrast (ECC) images were collected. Figure 3b presents a typical channeling-contrast image taken from the cross section of a 2h sample. As shown in the figure, the microstructure exhibits a lath morphology, with several grains being partitioned by laths of differing orientations. Closer examination reveals that the laths appear to be confined within single grains, which indicates a martensitic reaction that is consistent with previous reports by for this TRIP DP-HEA [45]. The growth of these laths is limited by grain boundaries (shown in red arrows) and these features appear to extend either partially through a grain (shown in blue arrows) or terminate at an opposing grain boundary or lath interface (shown in yellow arrows). A small number of secondary precipitates and casting defects (microporosity) appear as black spherical features under channeling-contrast imaging. EDX mapping (see Supplementary Material) identified the vast majority of these inclusions to be Mn-rich precipitates. Precipitates of similar morphologies are seen in ECC imaging of this HEA system in a study from Su et al. [49]. In order to correlate the observed XRD phases with features in the microstructure, EBSD analysis was performed on each of the specimens. Figure 4a contains an EBSD phase map of the 2h sample. As illustrated in the figure, the EBSD phase map identifies the parent austenite grain as the FCC phase and the martensitic laths as the HCP phase. These findings confirm the dual-phase microstructure detected in XRD analysis and align with reports from Li et al. [29] for this TRIP DP-HEA. Despite the detection of a small number of Mn-rich precipitates, EDX mapping of the bulk revealed good mixing and homogeneity of constituents in the dual-phase microstructure (Figure 4b). Quantitative chemical analysis showed that the constituent elements are close to the target molar composition (Cr = 9%, Mn = 29%, Co = 10%, Fe = bal.).

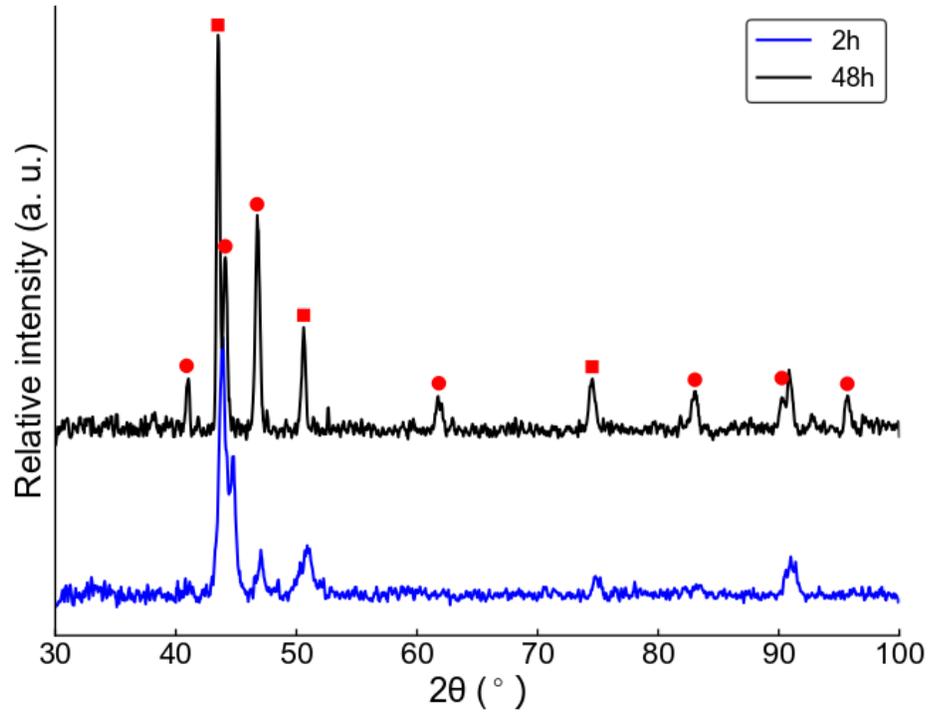

**Figure 2:** XRD data collected after heat treatment at 1200 ºC for 2 and 48 hours. Circular markers represent peaks belonging to the HCP phase and the square markers represent peaks corresponding to the FCC phase.

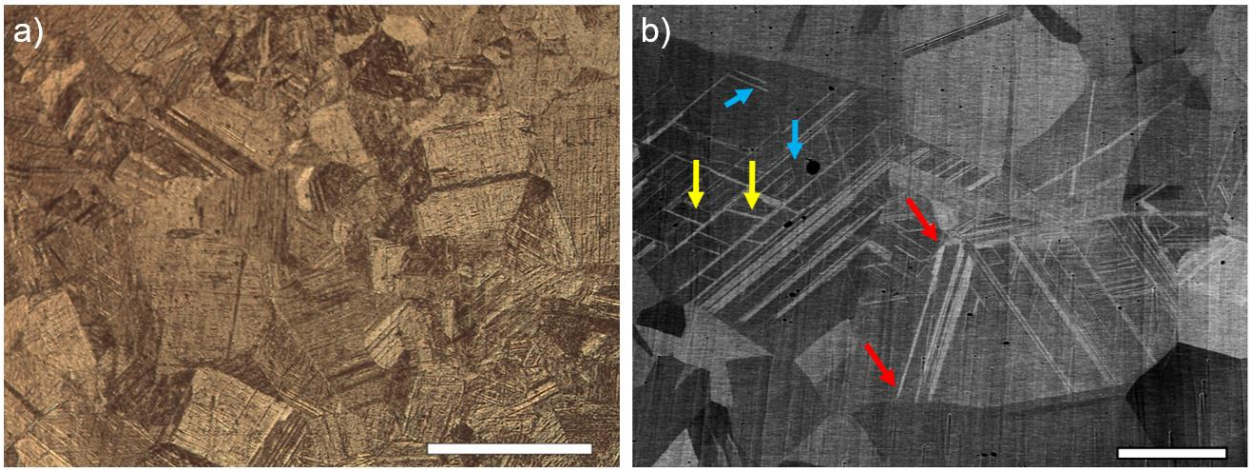

**Figure 3:** a) An optical micrograph of the 2h sample etched with CuCl$_2$ solution. b) A channeling-contrast image of the 2h sample. Several features of the lath structures are indicated with arrows (see main text for details). The scale bars represent lengths of 400 and 100 µm, respectively.

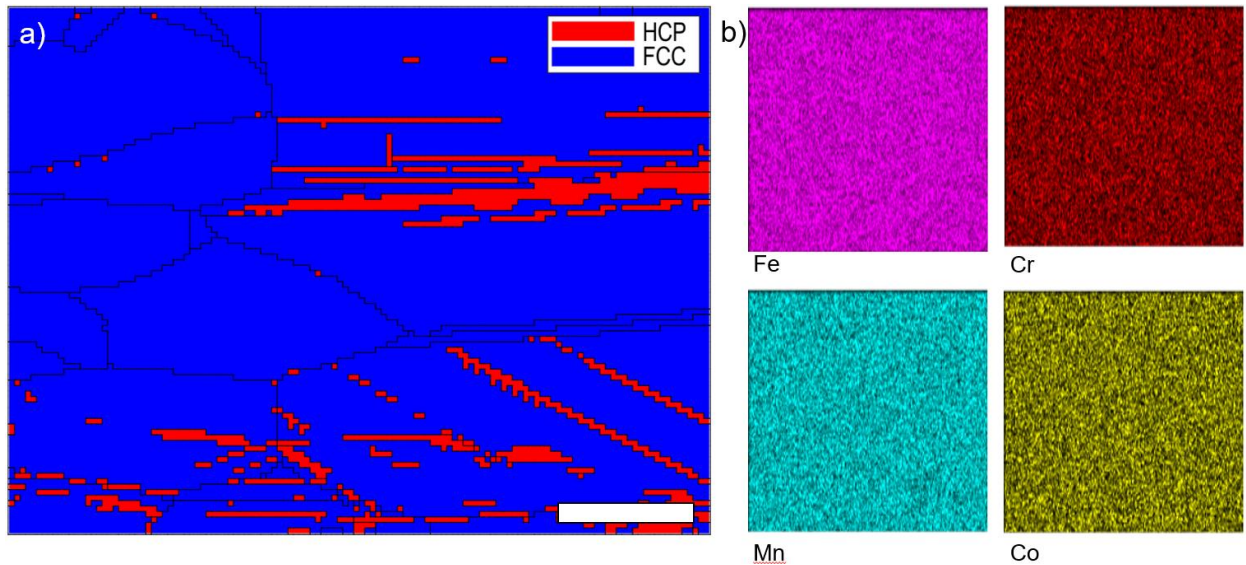

**Figure 4:** a) EBSD phase map of the 2h specimen. b) EDX maps showing the distribution of constituents in the microstructure. The scale bar in a) represents a length of 50 µm. The EDX maps in b) were collected over an area of 50 x 50 µm.

### 3.2. Mechanical behavior

Figure 5 presents representative engineering stress – engineering strain plots for the samples annealed at different holding times. The relevant mechanical parameters such as the yield strength ($\sigma_y$), ultimate tensile strength ($\sigma_u$), and total elongation have been extracted from the tensile curves and are presented in Table 1. The yield stress is calculated using the 0.2% offset method. As shown in Table 1, the 2h sample showed the highest elongation (33 ± 2 %), with $\sigma_y$ and $\sigma_u$ measured at 190 ± 6 and 523 ± 12 MPa, respectively. With the increase in holding time to 12h, $\sigma_u$ and elongation decreased significantly with respect to the 2h sample (17 and 21% respectively). Further increase in holding time to 48h, results in an additional decrease in tensile strength and elongation (28% and 73% respectively in comparison to the 2h sample). It is interesting to note that elongation decreases despite an increase in the grain size at longer heat treatment times. Examination of the post-yield portion of the mechanical testing data reveals that each sample exhibited similar work hardening behavior. Yet, longer heat treatment times resulted in earlier fracture events, which is consistent with the measured reductions in tensile strengths and elongation. We attribute this somewhat counterintuitive trend to the increased formation of Mn rich precipitates (see Supplementary Information) and to the reduction in volume fraction of FCC (ductile) phase at longer heat treatments. Casting porosity is not believed to significantly influence sample ductility due to the low statistical error in elongation and the clear link to heat treatment duration that emerges from mechanical testing data. To further explore changes in mechanical behavior with annealing time, we performed micro-indentation testing. This data is summarized along with the other mechanical properties in Table 1. Results indicate that the microhardness was highest for the 2h sample (277 ± 19 HV) and monotonically decreased to a low of 234 ± 12 HV for the 48h sample. As expected, the trend in indentation hardness measurements tracks well with changes $\sigma_u$, which is consistent with the study Zhang et al. [50] on the relationship of hardness and tensile strength.

Perhaps the most intriguing result from mechanical testing data is the absence of a change in yield strengths. Indeed, examination of $\sigma_y$ data in Table 1 shows a near constant yield stress over each heat treatment condition. To best visualize this result, we have plotted the collected microstructure and mechanical property measurements in Figure 6. It can be seen from the figure that with the increase in grain size from 102.1 ± 11.2 μm (2h sample) to 404.4 ± 23.2 μm (48h sample), the yield strength remains constant. This counterintuitive result is at odds with expectations for grain growth-related weakening, as reported by many authors for similar materials (i.e., austenitic steels) [44,51,52]. For instance, Xie et al. [53] studied size effects in the mechanical behavior of a Fe-Cr-Ni TRIP alloy under annealing treatments and reported an anticipated decrease in yield strength, which is consistent with the Hall-Petch effect. Yet, the DP-HEA presented herein is found to be surprisingly resilient to changes in parent crystal size. The physical reasons for this interesting behavior are examined in the discussion section. It should be noted that indentation hardness is not anticipated to capture the trends observed in the yield stress due to the dramatically different mechanics that underpin the flow conditions of each measurement. Brooks et al. [54] provides an excellent analysis that emphasizes the work hardening regime as a determining factor in hardness measurement.

**Table 1:** Microstructure and mechanical properties measured from DP-HEA samples.

| Sample | $\sigma_y$ (MPa) | $\sigma_u$ (MPa) | Elongation (%) | Hardness (HV) | Grain size (μm) | HCP (%) | Inter-lath spacing (μm) |
|---|---|---|---|---|---|---|---|
| 2h | 190 ± 6 | 523 ± 12 | 33 ± 2 | 277 ± 19 | 102.1 ± 11.2 | 9.1 ± 1.5 | 9.7 ± 2.1 |
| 12h | 185 ± 8 | 433 ± 6 | 26 ± 1 | 265 ± 15 | 170.7 ± 15.1 | 13.8 ± 1.0 | 7.9 ± 2.1 |
| 24h | 185 ± 2 | 453 ± 8 | 18 ± 1 | 250 ± 19 | 270.0 ± 19.0 | 18.2 ± 1.5 | 4.7 ± 1.8 |
| 48h | 190 ± 2 | 375 ± 3 | 9 ± 1 | 234 ± 12 | 404.4 ± 23.2 | 23.9 ± 0.5 | 4.2 ± 1.6 |

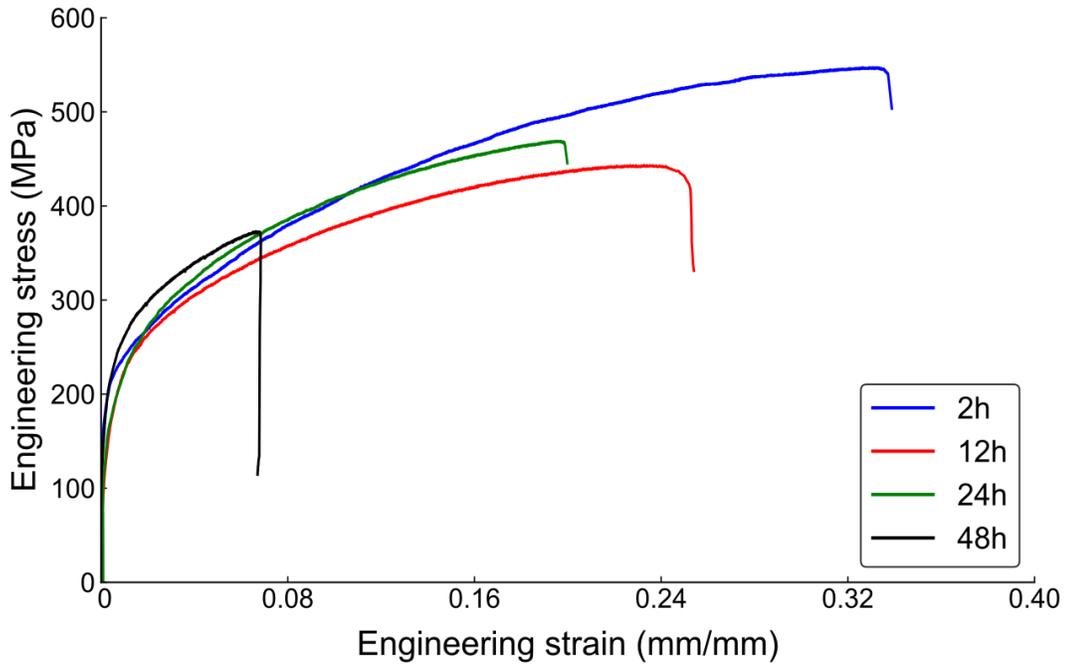

**Figure 5:** Representative tensile stress-strain curves of the DP-HEA samples heat treated for different holding times.

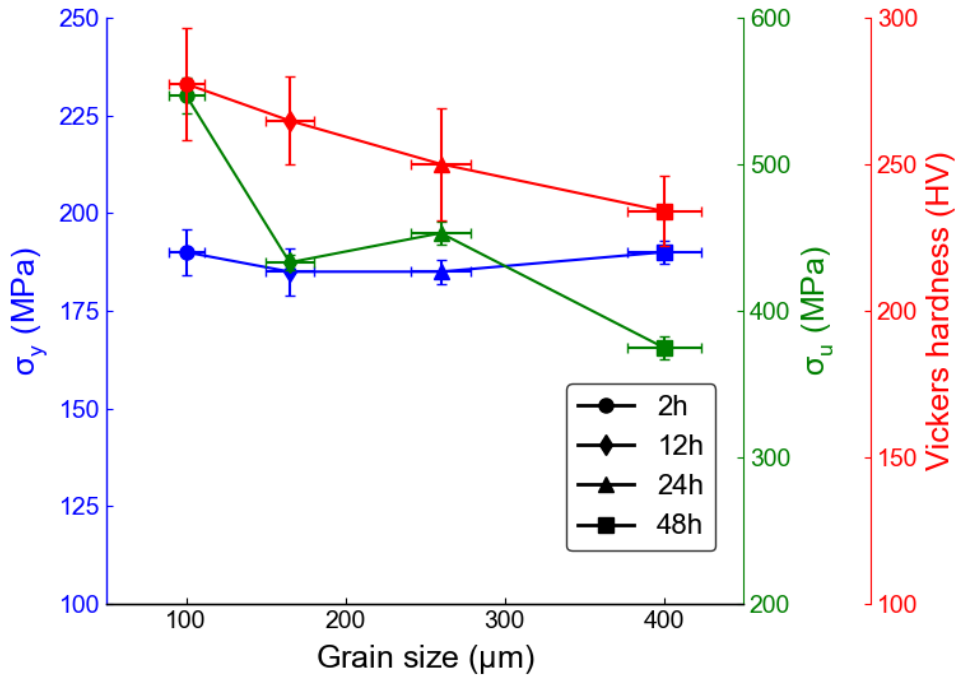

**Figure 6:** Effect of grain size on the yield strength, Vickers microhardness and ultimate tensile strength of the DP-HEA samples heat treated at different holding times.

## 4. DISCUSSION

In order to understand the insensitivity of the yield strength to grain growth, a more detailed analysis of the microstructure under each heat treatment condition has been pursued. EBSD phase maps for the 2h, 12h, 24h and 48h samples (Figure 7) reveal that the volume fraction of the HCP phase increases monotonically with heat treatment time. Table 1 contains the phase statistics for each heat treatment condition, which were measured from multiple EBSD maps at each heat treatment condition. The sample heat treated for 2h contains the lowest HCP phase fraction (9.1 ± 1.5 %) while the 48h sample exhibited the highest volume fraction (23.9 ± 0.5 %). The volume fractions of 12h and 24h samples were measured at 13.8 ± 1.0 and 18.2 ± 1.5 % respectively. The increase in volume fraction of the HCP phase can be attributed to a decrease in stability of the parent FCC matrix. Several experimental [10,39–41] and theoretical studies [44,55] demonstrate the increased susceptibility of larger grains to the martensitic transformation. One proposed mechanism is guided by nucleation of the HCP phase from stacking faults, which are present in an increased density in larger crystals [56]. An alternative mechanism is proposed by Takaki et al. [57], where the martensitic transformation is driven by lath branching, which we observed in greater frequency in the larger crystals of our specimens. While the collected results and relevant literature can certainly rationalize the observed changes in HCP volume fraction, a clear understanding for the insensitivity of the yield strength to grain growth-induced weakening effects requires further examination.

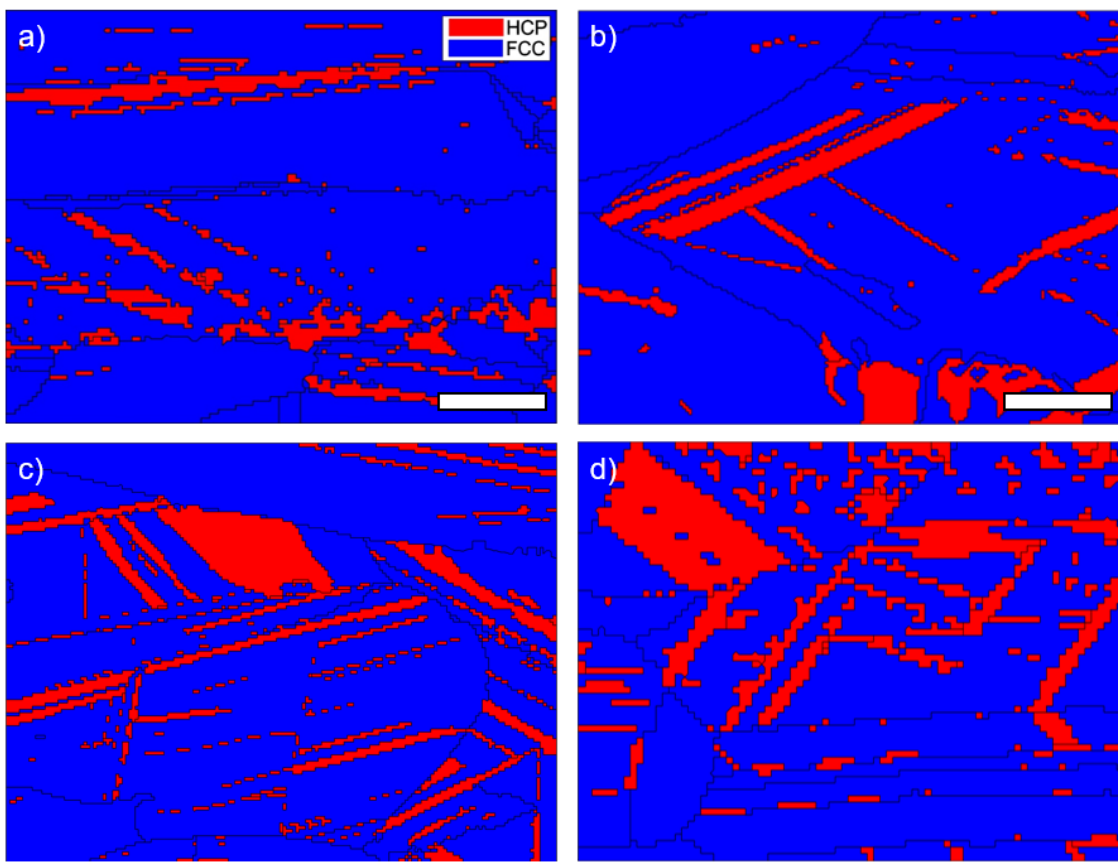

**Figure 7:** EBSD phase maps for samples annealed at 1200 °C for a) 2 hours, b) 12 hours, c) 24 hours, and d) 48 hours. The relative increase in the HCP volume fraction is clearly visible. The scale bar in a) represents a length of 50 μm. b)-d) are captured at the same magnification with a scale bar of 25 μm. Blue color in the phase map represents the FCC matrix and red denotes the HCP phase. The black lines represent grain and phase boundaries in the microstructure.

Examination of the HCP lath morphologies across the different annealing conditions yields a potential explanation. Namely, the evolution of the inter-lath spacing (i.e., distance between HCP laths) provides interesting insight. Figure 8a presents a comparison of DP-HEA grain size, inter-lath spacing, and volume fraction of HCP phase. Figure 8b and Figure 8c provides representative examples of channeling-contrast images in the 2h and 48h condition that are used to measure the inter-lath spacing. Similar images were taken to measure the inter-lath spacing for the 12h and 24h samples. Examination of the collected measurements reveals that the average inter-lath spacing reduced with increasing grain size. This finding is consistent with the trends in the HCP volume fraction, where increases in the HCP phase resulted in an increase in the lath density. The extremes of measurements are found in the 2h and 48h samples, where inter-lath spacings of $9.7 \pm 2.1$ μm and $4.2 \pm 1.6$ μm are measured, respectively. These findings are consistent with the available literature. Takaki et al. [57] reported that with the increase in parent FCC grain size, branching of the existing HCP laths increases, which results in crystal partitioning. The decrease in the inter-lath spacing can also rationalize the measured decrease in elongation, where flow is hindered by partitioning of the intracrystalline material and by introduction of the less ductile HCP phase. We interpret this lath-induced partitioning in the DP-HEA microstructure as being responsible for the retention of yield strength despite dramatic grain growth. That is, the lath boundaries act as barriers to dislocation motion that counteract grain boundary softening of the parent FCC phase. The importance of martensitic interfaces in acting as barriers to flow finds support in the steel literature [58–60] as well as in TRIP-enabled HEAs [2]. Du et al. [61] offers direct evidence of this austenite/martensite interfacial strengthening mechanism through a series of targeted uniaxial micro-tension experiments. Here, the authors performed micro-tensile tests on samples having martensite laths and directly observed the evidence of dislocation pinning at the interphase boundary. In a separate study, Chen et al. [62] observed dislocation pile-up phenomenon at lath boundaries under *in situ* transmission electron imaging, highlighting the role of these interfaces in facilitating strengthening. Within the context of the current work, these results offer an understanding for the perceived insensitivity of the DP-HEA to the grain growth-induced weakening effect. Namely, the increased susceptibility of the DP-HEA microstructure to martensitic formation under prolonged heat treatments restores size-related strengthening mechanisms, despite significant crystal growth in the parent FCC microstructure. We note that the parent grains studied herein are very large, which is a consequence of the intent of this study (i.e., examination of mechanical properties after prolonged, high temperature heat treatment). The accuracy of extrapolation of these findings to smaller parent grain sizes requires further study.

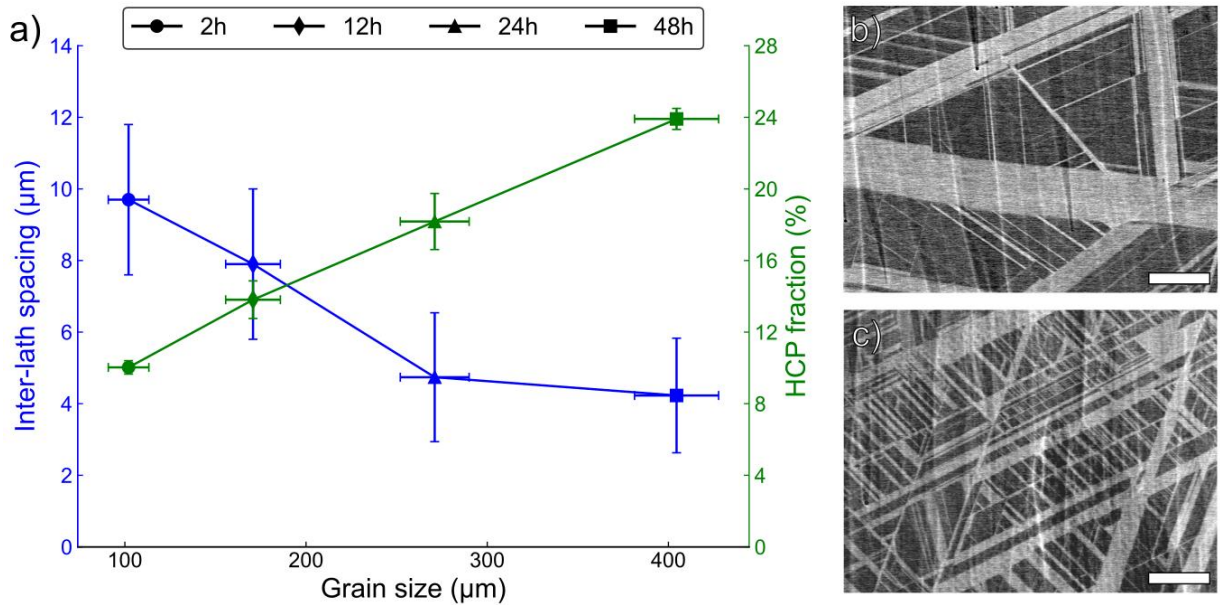

**Figure 8:** a) The average inter-lath spacing for each sample studied herein. These results are mapped to the average grain size of the FCC parent crystal and the HCP phase fraction (secondary axis). b) ECC images taken from the cross section of the b) 2h and c) 48h samples. The scale bars in b) and c) measure 10 μm. Comet-tailing features in ECC images are an artifact of ion milling.

To analyze the retention of yield strength over prolonged heat treatments, we calculate the contributions of grain growth-induced weakening and HCP lath-induced strengthening over the different annealing conditions using a two-parameter Hall-Petch model. This mechanistic model is based on the approach of Wang et al. [63], and has been refined from the work of Su et al. [49]. Here, the normalized changes in the contributions of lath boundary ($\Delta\bar{\sigma}_{lb}$) and grain boundary ($\Delta\bar{\sigma}_{gb}$) mechanisms to the normalized yield strength ($\bar{\sigma}_y$) are calculated as:

$$\bar{\sigma}_y = 1 + f_{lb}\Delta\bar{\sigma}_{lb} + f_{gb}\Delta\bar{\sigma}_{gb} \tag{1}$$

where $f_{lb}$ and $f_{gb}$ are the volume fractions of grains with HCP laths and grains undergoing growth during heat treatment (taken as unity), respectively. The volume fraction of grains containing HCP laths is estimated using the number fractions measured from low magnification SEM images. $f_{lb}$ of 0.2, 0.33, 0.81, and 1, are measured for the 2h, 12h, 24h, and 48h samples, respectively. Each of the strength parameters are normalized using the yield strength of the 2h sample (e.g., $\bar{\sigma}_y = \sigma_y/\sigma_{y,2h}$). The contributions of grain boundaries and HCP-laths to weakening/strengthening of the microstructure are calculated as:

$$\Delta\bar{\sigma}_{gb} = \frac{k_{gb}}{\sigma_{y,2h}}\left(\frac{1}{\sqrt{d}} - \frac{1}{\sqrt{d_{2h}}}\right) \tag{2a}$$

$$\Delta\bar{\sigma}_{lb} = \frac{k_{lb}}{\sigma_{y,2h}}\left(\frac{1}{\sqrt{t}} - \frac{1}{\sqrt{t_{2h}}}\right) \tag{2b}$$

where $d$ is the grain size, $t$ is the inter-lath spacing, and $k_{gb}$ and $k_{lb}$ are the Hall-Petch coefficients of the grain boundary and lath boundary mechanisms, respectively. The values of $k_{gb}$ and $k_{lb}$ are taken from the published literature for a similar system and set as 573 MPa.$\mu m^{1/2}$ and 195 MPa.$\mu m^{1/2}$, respectively [49]. It should be noted that in this reference the lath boundary coefficient is determined for twin lamella, but we find good agreement with the data presented herein. This may be due to the HCP configuration of the coherent twin boundary in FCC materials.

Results of this mechanistic modeling effort are presented in Figure 9. As shown in the figure, predictions for the normalized yield strengths match well with experimental measurements. It should be noted that no fitting procedures were used in modeling. All inputs were measured from microscopy images and the Hall-Petch parameters (i.e., $k_{gb}$ and $k_{lb}$) were obtained from the literature. In the extreme condition (48h sample), the grain boundary weakening led to a 15% drop in the yield strength and the HCP-lath induced hardening increased yield strength by 17% relative to the 2h reference. The absolute contributions of these values to the yield strength cannot be determined without dedicated single crystal measurements.

This analysis demonstrates the balancing of weakening and strengthening mechanisms that lead to the retention of yield strength at larger parent grain sizes for this TRIP DP-HEA. It is noteworthy that a review of the existing TRIP steel literature shows that this effect appears to be unique. That is, TRIP-enabled steels do indeed exhibit conventional grain growth-induced weakening at parent grain sizes in the ~ 100 μm range, comparable to those studied herein [7]. These findings motivate future work to investigate the unique microstructure evolution phenomena that underpin the insensitivity of this TRIP DP-HEA to grain growth-induced weakening. Furthermore, as HEAs are considered promising candidates for high temperature applications, the mechanical resiliency of this DP-HEA microstructure to aggressive heat treatments offers new opportunities for high performance structural alloy design.

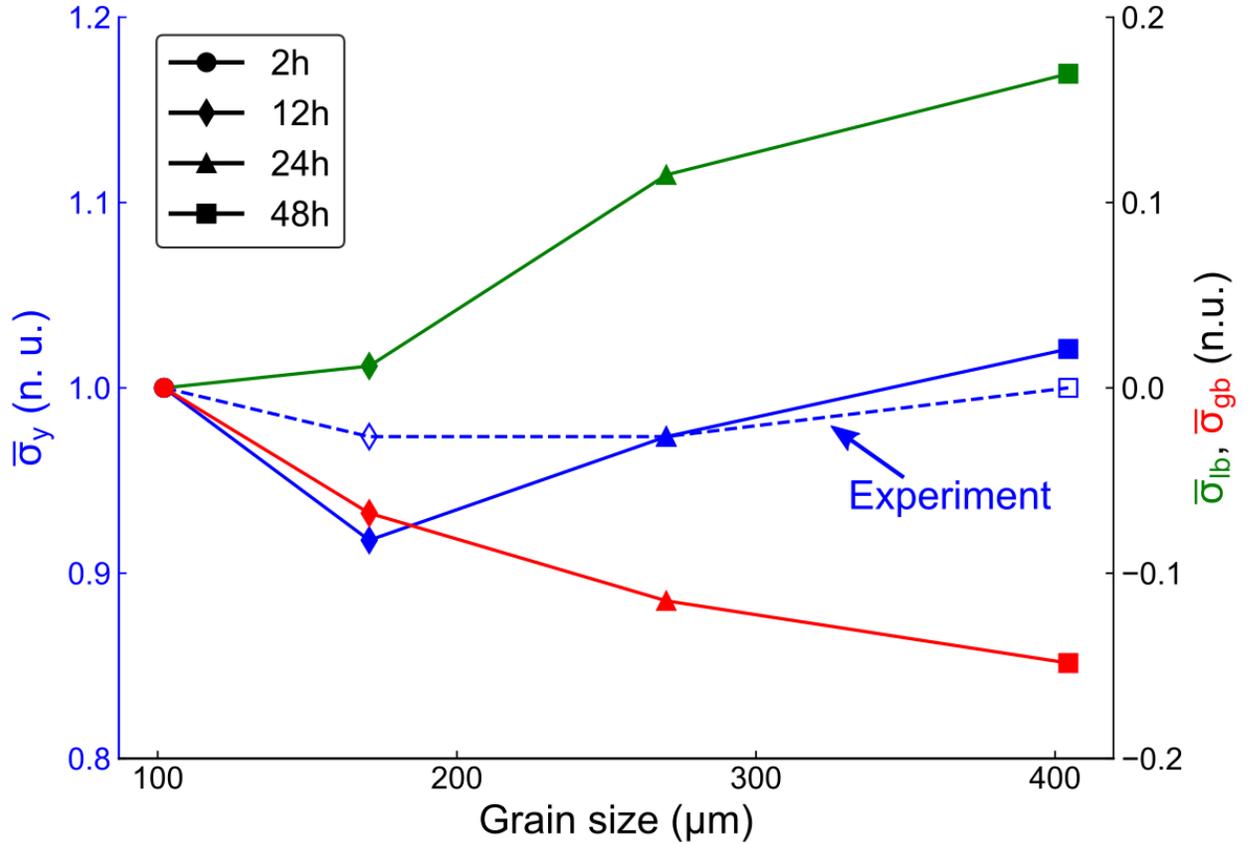

**Figure 9:** Estimates of the relative contributions of grain boundary and HCP lath boundary mechanisms to the yield strength over prolonged heat treatment. All data are normalized to the yield strength of the 2h sample. Overall model predictions of yield strength are provided in solid blue stroke with filled markers. The experimental measurements from tensile testing are provided in dashed blue stroke with open markers. $\Delta\bar{\sigma}_{lb}$ and $\Delta\bar{\sigma}_{gb}$ are read from the secondary y-axis and the datapoints match the color of their secondary y-axis labels.

## 5. CONCLUSIONS

The effect of prolonged heat treatments on the evolution of microstructure and mechanical properties in a TRIP DP-HEA has been studied. XRD and EBSD analysis confirmed the retention of a duplex FCC/HCP microstructure for all heat treatment conditions. Optical and electron microscopy revealed a lath microstructure that is characteristic of this TRIP system, where the FCC and HCP phases are present as the austenite and martensite, respectively. The length of heat treatment was found to be positively correlated with the HCP phase fraction and lath density in the DP-HEA microstructure. Samples heat treated for the shortest durations (2 hours) were found to possess the largest fraction of retained FCC austenite. Microstructure analysis revealed a monotonic increase in the parent FCC grain size with heat treatment duration, where the 2h sample exhibited the smallest FCC grain size (102.1 ± 11.2 μm) and 48h sample exhibited largest FCC grain size (404.4 ± 23.2 μm). Uniaxial tensile testing revealed an anomalous stability in the yield strength of DP-HEA samples over all heat treatment durations, despite a near four-fold increase in the parent crystal size. Further microstructure analysis revealed the role of the inter-lath spacing in underpinning this observed resiliency to grain growth-induced weakening. Namely, the increase in HCP phase fraction was found to result in an increased lath number density within the DP-HEA microstructure, which led to a significant partitioning of the parent FCC grains. Thus, the HCP lath boundaries provide additional obstacles to dislocation motion along with parent FCC grain boundaries, which compensated for grain growth-induced weakening. A two-parameter model was implemented to calculate the relative contributions of weakening and strengthening mechanisms to the yield stress which showed excellent agreement with experimental measurements. The results of this work demonstrate a promising resiliency of this DP-HEA microstructure, which opens opportunities for applications in extreme environments.

## ACKNOWLEDGEMENTS

This work was supported by funding from the University of Illinois at Chicago. This work made use of instruments in the Electron Microscopy Core of UIC's Research Resources Center.

## AUTHOR CONTRIBUTIONS

J. A. performed the alloy synthesis, experiments, and analysis under guidance from M. D. Both authors co-wrote the manuscript. M. D. conceived the study and directed the effort.

## ADDITIONAL INFORMATION

The supplementary material is available online from the publisher or by request from the corresponding author (mattdaly@uic.edu).

## DATA AVAILABILITY

The raw/processed data required to reproduce these findings cannot be shared at this time due to technical or time limitations.